\documentclass[conference]{IEEEtran}
\IEEEoverridecommandlockouts
\usepackage{cite}
\usepackage{amsmath,amssymb,amsfonts}
\usepackage{algorithmic}
\usepackage{graphicx}
\usepackage{textcomp}
\usepackage{xcolor}
\usepackage{url}
\usepackage{multirow}
\def\BibTeX{{\rm B\kern-.05em{\sc i\kern-.025em b}\kern-.08em
    T\kern-.1667em\lower.7ex\hbox{E}\kern-.125emX}}
\begin{document}
\vbadness=10000
\title{Emotional Styles Hide in Deep Speaker Embeddings: Disentangle Deep Speaker Embeddings for Speaker Clustering
}

\author{\IEEEauthorblockN{Chaohao Lin}
\IEEEauthorblockA{\textit{Electrical and Computer Engineering} \\
\textit{Florida International University}\\
Miami, USA \\
clin027@fiu.edu}
\and
\IEEEauthorblockN{Xu Zheng}
\IEEEauthorblockA{\textit{Knight Foundation School of Computing and Information Sciences} \\
\textit{Florida International University}\\
Miami, USA \\
xzhen019@fiu.edu}
\and
\IEEEauthorblockN{Kaida Wu}
\IEEEauthorblockA{\textit{Electrical and Computer Engineering} \\
\textit{Florida International University}\\
Miami, USA \\
kwu020@fiu.edu}
\and
\IEEEauthorblockN{Peihao Xiang}
\IEEEauthorblockA{\textit{Electrical and Computer Engineering} \\
\textit{Florida International University}\\
Miami, USA \\
pxian001@fiu.edu}
\and
\IEEEauthorblockN{Ou Bai}
\IEEEauthorblockA{\textit{Electrical and Computer Engineering} \\
\textit{Florida International University}\\
Miami, USA \\
obai@fiu.edu}

}

\maketitle

\begin{abstract}
Speaker clustering is the task of identifying the unique speakers in a set of audio recordings (each belonging to exactly one speaker) without knowing who and how many speakers are present in the entire data, which is essential for speaker diarization processes. Recently, off-the-shelf deep speaker embedding models have been leveraged to capture speaker characteristics. However, speeches containing emotional expressions pose significant challenges, often affecting the accuracy of speaker embeddings and leading to a decline in speaker clustering performance. To tackle this problem, we propose DTG-VAE, a novel disentanglement method that enhances clustering within a Variational Autoencoder (VAE) framework. This study reveals a direct link between emotional states and the effectiveness of deep speaker embeddings. As demonstrated in our experiments, DTG-VAE extracts more robust speaker embeddings and significantly enhances speaker clustering performance. Our code is available: \url{https://github.com/Toby28/DDSESC}.
\end{abstract}


\section{Introduction}
Speaker clustering is the process of grouping audio segments based on the identity of the speaker, without prior knowledge of who the speakers are or how many there are. It is widely used in tasks like speaker diarization and voice-based analytics \cite{shum2013unsupervised, yu2017active, park2022review}.

Researchers have developed various pretrained deep speaker embedding models to extract speaker characteristics from speech, including \texttt{d-vector} \cite{wan2018generalized}, \texttt{r-vector} \cite{villalba2020statervector} , and \texttt{ECAPA-TDNN} \cite{desplanques2020ecapa}. Leveraging these off-the-shelf models for speaker clustering has become a common approach \cite{ulgen2024revealing,tong2022graph,wang2024semi,zeng2023sef,xu2021speaker}. While effective, these models often struggle to distinguish emotional styles when extracting deep speaker embeddings.

However, in everyday interactions, speech is infused with a wide range of emotions and tones—such as happiness, sadness, and anger—that enrich communication and self-expression. This negligence can pose challenges in accurately identifying and distinguishing speakers when emotions alter vocal characteristics.

While variations in emotional style can hinder speaker clustering when using pretrained deep speaker embedding models, relatively little research has explored how to disentangle emotional styles from these embeddings \cite{pappagari2020x}. Nevertheless, previous research has observed that speech emotion and content can negatively affect speaker characteristic representations. To address this issue, various methods have been proposed to disentangle emotional style from speech signals \cite{brima2023learning, lu2023speechtriplenet, tu2024contrastive}. Although these approaches work well, these approaches require training speaker embedding models from scratch rather than utilizing off-the-shelf deep speaker embedding models, making adaptation to different speakers more challenging. On the other hand, some researchers have focused on converting emotional speech into neutral speech to extract more accurate deep speaker embeddings and reduce the impact of emotional styles \cite{li2023identity,li2023sec,oh2024durflex,rizos2020stargan,ghosh2023emo, lian2022robust}. These aspects have not fundamentally solved the deep speaker embedding problem. Ulgen et al. challenge the traditional assumption that emotional information is absent from speaker embeddings. They propose that deep speaker embeddings, in fact, contain valuable emotion-related information \cite{ulgen2024revealing, li2017deep}.

\begin{table*}[t]
\centering
\caption{Comparative analysis of pre-trained deep speaker embedding models for speaker clustering using neutral and emotional speech samples from the ESD and IEMOCAP Databases. Results are presented in terms of the performance of neutral speech / emotional speech. Drop${\Delta}$ refers to the reduction in accuracy when transitioning from neutral to emotional speech.}
\scalebox{1.0}{
\setlength{\tabcolsep}{12pt}
\begin{tabular}{c|c|cc|cc|cc}
\hline
\multirow{2}{*}{} & \multirow{2}{*}{} & \multicolumn{6}{|c}{Neutral Speech/Emotional Speech} \\
\cline{3-8}
 &  & NMI$ [0,1]\uparrow $ & Drop${\Delta}$ & ARI$[0,1]\uparrow$ & Drop${\Delta}$ & Silhouette$[-1,1]\uparrow$ & Drop${\Delta}$ \\
\hline
& & \multicolumn{6}{c}{ESD}\\
\hline
\multirow{3}{*}{d-vector} & KM & 0.92/0.81 & 0.11 & 0.84/0.73 & 0.11 & 0.21/0.13 & 0.08 \\
& SC & 0.94/0.93 & 0.01 & 0.86/0.87 & -0.01 & 0.22/0.14 & 0.08 \\
& AC & 0.96/0.90 & 0.06 & 0.96/0.85 & 0.11 & 0.21/0.13 & 0.08 \\
\hline

\multirow{3}{*}{r-vector} & KM & 0.95/0.90 & 0.05 & 0.87/0.87 & 0.00 & 0.24/0.20 & 0.04 \\
& SC & 0.95/0.95 & 0.00 & 0.87/0.87 & 0.00 & 0.25/0.18 & 0.07 \\
& AC & 1.00/0.99 & 0.01 & 1.00/0.99 & 0.01 & 0.30/0.21 & 0.09 \\
\hline

\multirow{3}{*}{ECAPA-TDNN} & KM & 1.00/0.95 & 0.05 & 1.00/0.87 & 0.13 & 0.28/0.16 & 0.12 \\
& SC & 0.99/0.93 & 0.06 & 0.99/0.82 & 0.17 & 0.19/0.18 & 0.01 \\
& AC & 0.99/0.99 & 0.00 & 0.99/0.99 & 0.00 & 0.27/0.19 & 0.08 \\
\hline
& & \multicolumn{6}{c}{IEMOCAP}\\
\hline
\multirow{3}{*}{d-vector} & KM & 0.46/0.29 & 0.17 & 0.34/0.16 & 0.18 & 0.09/0.07 & 0.02\\
& SC & 0.55/0.42 & 0.13 & 0.43/0.24 & 0.19 & 0.08/0.07 & 0.01\\
& AC & 0.46/0.31 & 0.15 & 0.32/0.18 & 0.14 & 0.06/0.04 & 0.02\\
\hline

\multirow{3}{*}{r-vector} & KM & 0.85/0.79 & 0.06 & 0.79/0.70 & 0.09 & 0.12/0.10 & 0.02\\
& SC & 0.89/0.87 & 0.02 & 0.88/0.85 & 0.03 & 0.14/0.10 & 0.04\\
& AC & 0.83/0.83 & 0.00 & 0.75/0.74 & 0.01 & 0.13/0.10 & 0.03\\
\hline

\multirow{3}{*}{ECAPA-TDNN} & KM & 0.80/0.79 & 0.01 & 0.74/0.67 & 0.07 & 0.12/0.09 & 0.03\\
& SC & 0.85/0.82 & 0.03 & 0.84/0.78 & 0.06 & 0.12/0.09 & 0.03\\
& AC & 0.77/0.76 & 0.01 & 0.69/0.62 & 0.07 & 0.11/0.08 & 0.03\\
\hline
\end{tabular}
}
\label{table1}
\end{table*}\begin{figure*}[t]
  \begin{minipage}[b]{.48\linewidth}
  \centerline{\includegraphics[width=8cm]{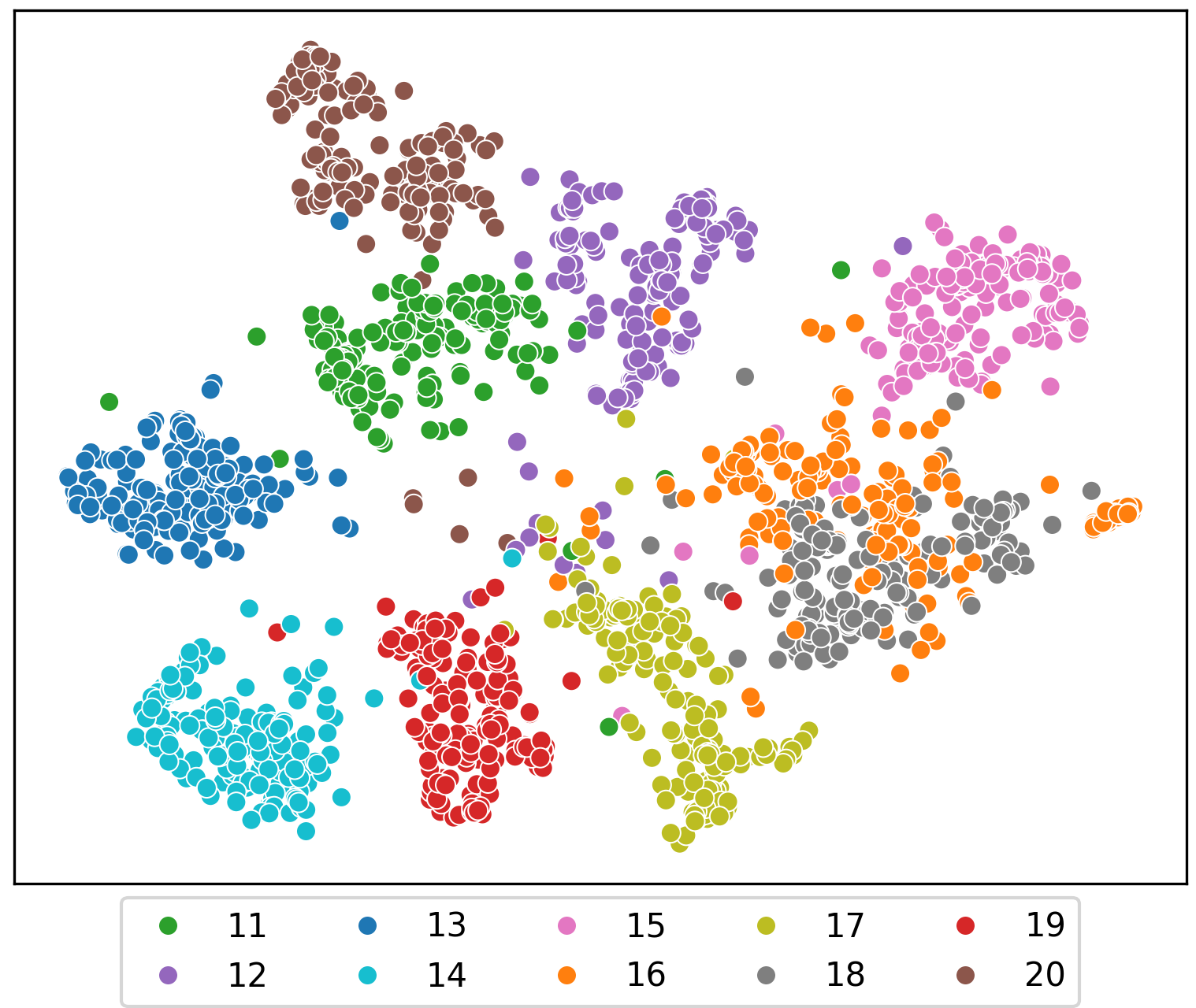}}
  \end{minipage}
\hfill
  \begin{minipage}[b]{0.48\linewidth}
  \centerline{\includegraphics[width=8cm]{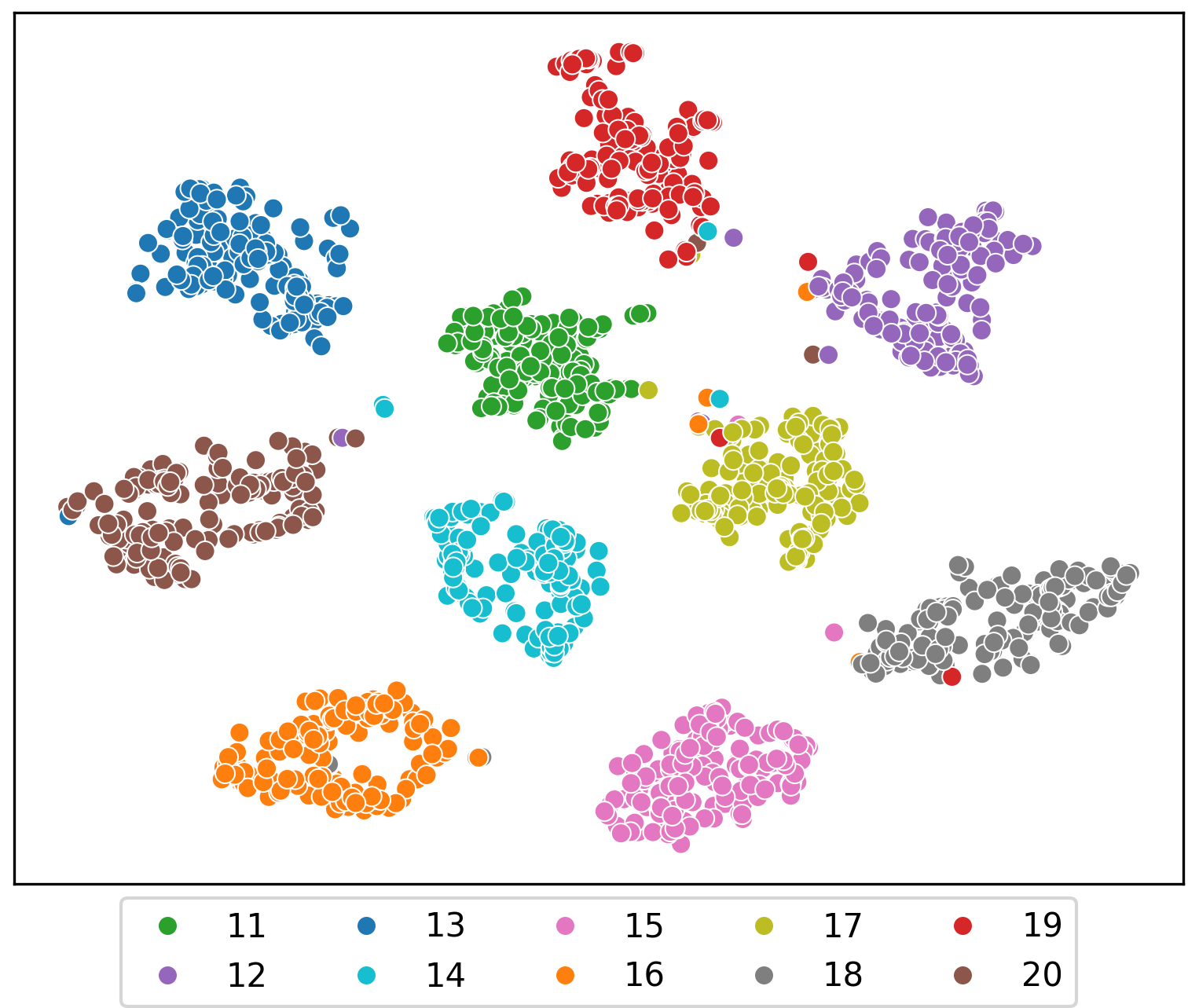}}
  \end{minipage}
\caption{T-SNE visualization of deep speaker embeddings for clustering speakers in the ESD dataset. The left panel shows the original performance, while the right panel displays the results of our approach.}
\label{fig1}
\end{figure*}

In prior studies, VAE-based frameworks are commonly utilized to disentangle spoken content and speaking style information from speech waveform and enhance speaker recognition accuracy \cite{tjandra2020unsupervised,lian2022towards,du2021disentanglement, lu2023speechtriplenet}. In this paper, we introduce a novel VAE-based fine-tuning framework that is designed to disentangle deep speaker embeddings for speaker clustering. This approach draws inspiration from prior research and adapts it to target the disentanglement of speech features specifically \cite{du2021disentanglement,lu2023speechtriplenet, wang2019vae}. Our method utilizes pre-trained deep speaker embedding models, eliminating the need for retraining. This approach not only extracts more effective deep speaker embeddings while mitigating the impact of emotional styles but also surpasses direct fine-tuning methods and other disentanglement techniques in speaker clustering tasks. The key contributions of this work can be summarized as follows:
\begin{itemize}
    \item We reveal a previously underappreciated problem: off-the-shelf pre-trained deep speaker embedding models still encode emotional styles, which can adversely affect speaker clustering performance—an aspect that existing research has largely overlooked.
    \item We introduce DTG-VAE, a novel VAE-based fine-tuning framework that accurately extracts and disentangles deep speaker embeddings from emotional style influences, resulting in enhanced speaker clustering performance.
    \item Our experimental results show that leveraging disentangled deep speaker embeddings enhances speaker clustering performance more effectively than both traditional fine-tuning methods and other VAE-based frameworks.
\end{itemize}

\section{Emotional Styles Hide in Deep Speaker Embeddings}
\label{sec:emotional}

\begin{figure*}[t]
  \begin{minipage}[b]{0.49\linewidth}
  \centerline{\includegraphics[width=9cm]{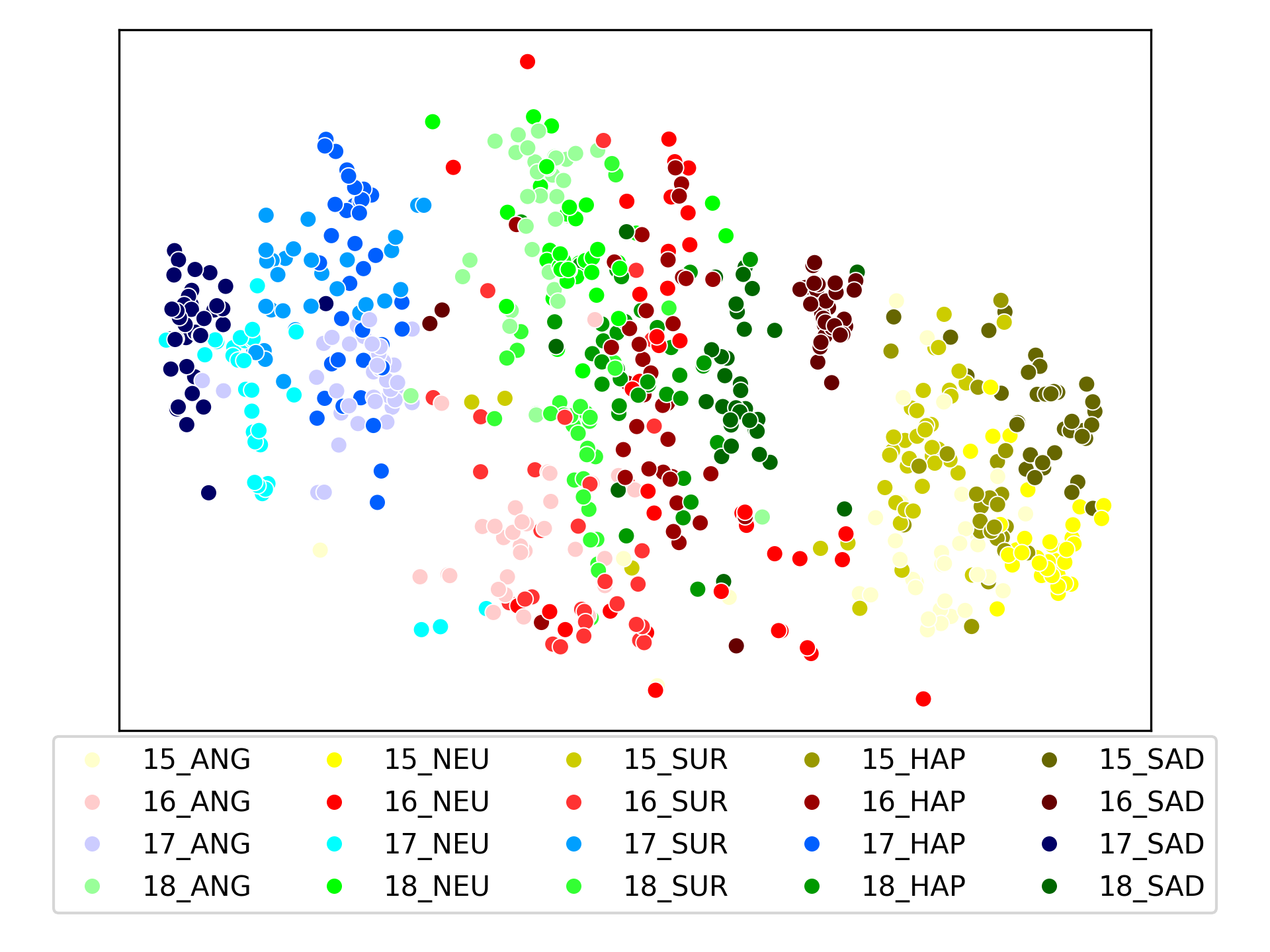}}
  \end{minipage}
\hfill
  \begin{minipage}[b]{0.49\linewidth}
  \centerline{\includegraphics[width=9cm]{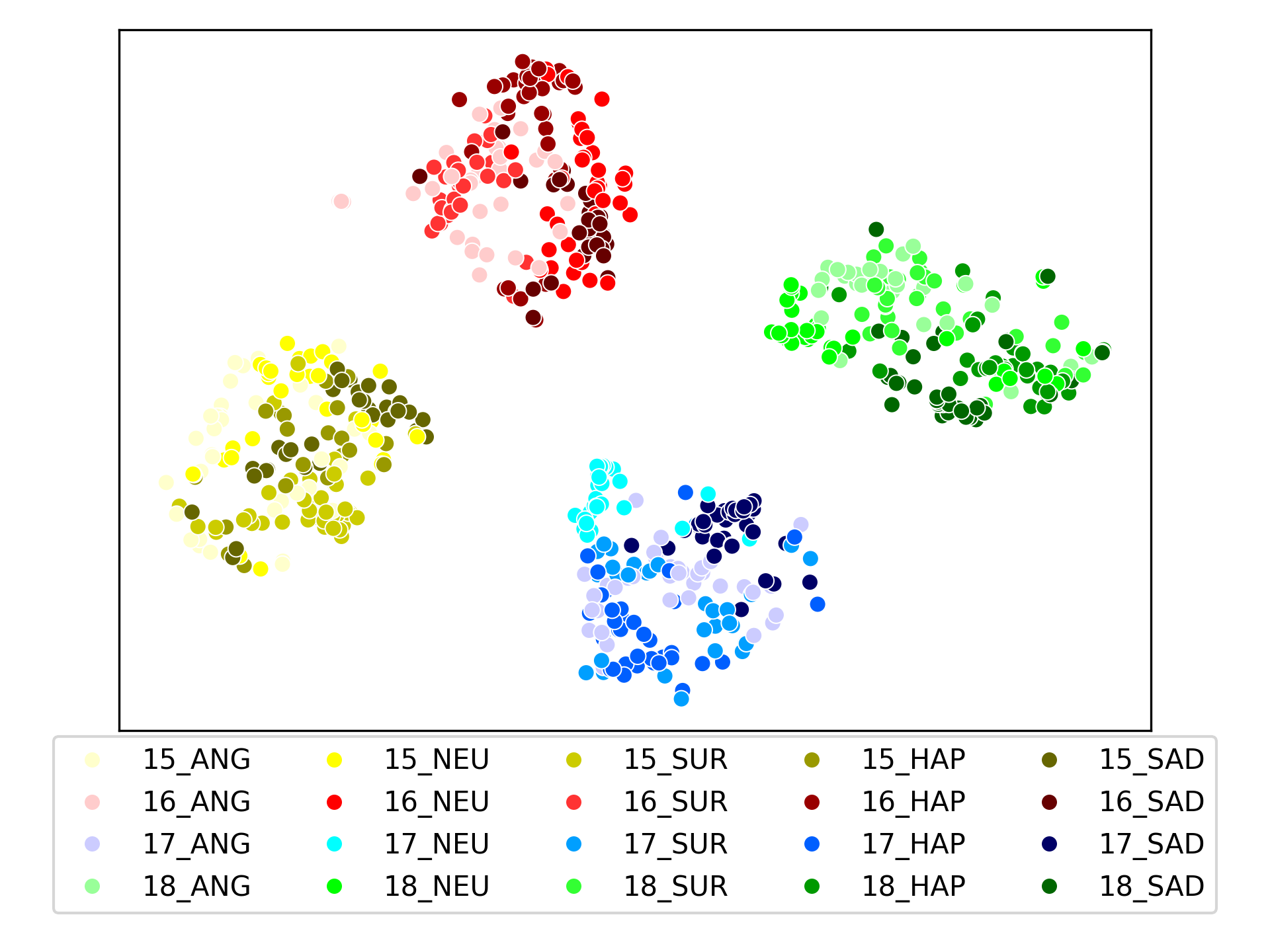}}
  \end{minipage}
\caption{A detailed visualization of Figure 1. Colors indicate the combination $\{speaker\_id\}\_\{emotion\}$, with the left panel representing the original and the right panel showcasing our approach.}
\label{fig2}
\end{figure*}

This section demonstrates that deep speaker embeddings inherently capture emotional styles, which in turn degrade the performance of speaker clustering tasks.
\subsection{Dataset and deep speaker embeddings}


To demonstrate that emotional styles are embedded in off-the-shelf deep speaker embeddings, we selected two widely recognized datasets: IEMOCAP\cite{busso2008iemocap} and ESD\cite{zhou2022emotional}. We focused our analysis on five emotional states—anger, neutral, surprise, happiness, and sadness. Speech samples expressing neutral emotion were isolated into a dedicated dataset, while those corresponding to the other four emotional states were combined into what we refer to as the Emotional dataset. We employed three off-the-shelf deep speaker embedding models, the d-vector\cite{wan2018generalized} and the r-vector\cite{villalba2020statervector}, as well as the ECAPA-TDNN\cite{desplanques2020ecapa}; all were pretrained on the VoxCeleb1 and VoxCeleb2 datasets\cite{nagrani2017voxceleb, chung2018voxceleb2}. These models have consistently demonstrated exceptional performance and robustness in speaker recognition tasks, particularly on the cleaned version of the VoxCeleb1 test set\cite{villalba2020statervector, wang2018speaker, zeinali2019but}.

\subsection{Speaker clustering and Evaluations}

We apply three clustering algorithms to the speaker clustering task: K-Means (KM), Spectral Clustering (SC), and Agglomerative Clustering (AC). We evaluate the performance of these methods using three established metrics: Normalized Mutual Information (NMI)\cite{vinh2009information}, Adjusted Rand Index (ARI)\cite{vinh2009information}, and the Silhouette Score\cite{rousseeuw1987silhouettes}.

Table \ref{table1} presents the speaker clustering results, revealing that the emotional versions of both the ESD and IEMOCAP datasets yield lower performance metrics compared to their neutral counterparts. Figure \ref{fig1} (left) displays a t-SNE visualization of speaker clusters derived from r-vector representations.

In particular, the d-vector speaker embedding exhibits a decline in performance across all three clustering algorithms when applied to emotional speech. On average, the following reductions were observed in key clustering metrics for the two datasets. For the ESD dataset, the Normalized Mutual Information (NMI) decreased by 0.06, the Adjusted Rand Index (ARI) dropped by 0.07, and the Silhouette Score fell by 0.08. In contrast, the IEMOCAP dataset experienced larger reductions, with NMI decreasing by 0.15, ARI by 0.17, and the Silhouette Score by 0.02. Although the r-vector and ECAPA-TDNN speaker embeddings generally show higher performance metrics overall, emotional speech consistently underperforms relative to neutral speech.

\section{Disentangle Deep Speaker Embeddings}
\label{sec:method}

\begin{figure}[ht]

\begin{minipage}[b]{1.0\linewidth}
  \centering
  \centerline{\includegraphics[width=8cm]{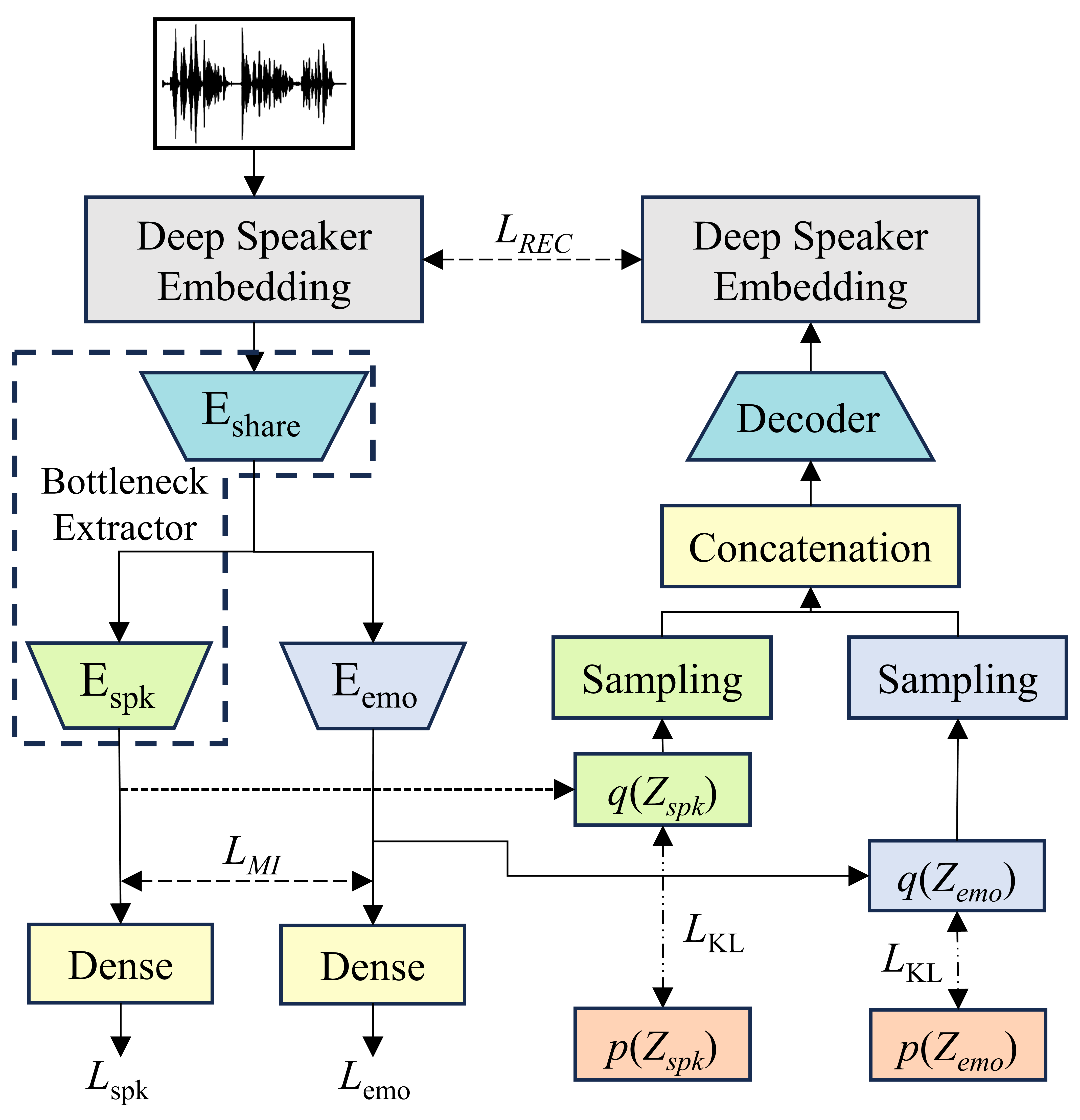}}
\end{minipage}
\caption{DTG-VAE Framework Training Process.}
\label{fig3}
\end{figure}

In this study, we propose a novel fine-tuning framework that disentangles emotional style and speaker identity within deep speaker embeddings. Figure \ref{fig3} illustrates the architecture of our model. We employ pre-trained deep speaker embedding models to convert raw speech waveforms into embeddings, denoted by
\begin{align}
  X=\{X^i\in \mathbb{R}^D|i=1,2,...,N\}
\end{align}
Here, \(N\) represents the number of speech utterances, while \(D\) denotes the number of dimensions in the vector representations generated by deep speaker embedding models. Our objective is to decompose \(X\) into two distinct latent representations—one capturing speaker identity and the other encoding emotional expression.


\subsection{Encoder-Decoder} 
Our encoder comprises three neural network modules: a shared encoder  $E_\text{share}$, a speaker encoder $E_\text{spk}$, and an emotion encoder $E_\text{emo}$. In the encoder stage, the shared encoder extracts features from the input data $X$. The speaker and emotion encoders, which share the same architecture, are designed to model the posterior distributions of speaker identity and emotion latent representations, respectively denoted as $q(Z_\text{spk}|X; \theta_\text{spk})$ and $q(Z_\text{emo}|X; \theta_\text{emo})$, where $\theta_\text{spk}$ and $\theta_\text{emo}$ represent the trainable parameters of the encoders. We characterize the posterior distributions for emotional style and speaker identity as multi dimensional isotropic Gaussian distributions, denoted by $q(Z_\text{spk}|X; \theta_\text{spk})=\mathcal{N}(Z_\text{spk}|\mu(X;\theta_\text{spk}),\sigma(X;\theta_\text{spk}))$ and $q(Z_\text{emo}|X; \theta_\text{emo})=\mathcal{N}(Z_\text{emo}|\mu(X;\theta_\text{emo}),\sigma(X;\theta_\text{emo}))$. 

Here $\mu(\cdot)$ and $\sigma(\cdot)$ are the function for mean and standard deviation calculation. We follow the VAE \cite{kingma2013auto} to use multi dimensional normal distributions as the prior knowledge of latent variables $Z_\text{spk}$ and $Z_\text{emo}$. Formally, we have $p(Z_\text{spk})=\mathcal{N}(Z_\text{spk}|0, I)$ and $p(Z_\text{emo})=\mathcal{N}(Z_\text{emo}|0, I)$.

Our decoder combines the speaker latent variable \(Z_\text{spk}\) and the emotion latent variable \(Z_\text{emo}\) to model the conditional probability distribution $q(X|Z_\text{spk},Z_\text{emo}; \theta_\text{dec})$ , where $\theta_\text{dec}$ denotes the trainable parameters in the decoder. This process essentially reconstructs the original deep speaker embeddings from the two latent variables.

\subsection{Training Objectives}
During training, our framework generates fixed-length deep speaker embeddings and optimizes them using five objectives.

\subsubsection{Reconstruction Loss}
Reconstruction loss focuses on maintaining data fidelity, which measures how well the decoded outputs match the original inputs. The better the reconstruction, the more effectively the VAE has learned the essential features of the data \cite{kingma2019introduction}.
\begin{align}
  \mathcal{L}_{\text{REC}} &= \frac{1}{2}\Vert X-\hat{X} \Vert^2_1 + \frac{1}{2}\Vert X-\hat{X} \Vert^2_2
\end{align}
where \(\hat{X}\) denotes reconstruction deep speaker embedding.
\subsubsection{KL divergence Loss}
The KL divergence acts as a regularizer by ensuring that the learned distribution of latent variables is close to the prior distribution, which helps balance accuracy and generalization.

\begin{equation}
    \begin{aligned}
        \mathcal{L}_{\text{KL}} &= \mathbb{E}_{p(X)}[KLD({q(Z_\text{spk}|X;\theta_\text{spk})||p(Z_\text{spk})})] \\
    &+ \mathbb{E}_{p(X)}[KLD({q(Z_\text{emo}|X;\theta_\text{emo})||p(Z_\text{emo})})]   
    \end{aligned}
    \label{eq:loss:kl}
\end{equation}
Where $KLD(\cdot)$ means the KL divergence calculation function.

\subsubsection{Decoupling loss}
The speaker's latent space should contain minimal emotional information. To achieve this, we propose a decoupling loss that encourages the speaker encoder and emotion encoder to capture distinct aspects of the data. Formally, the decoupling loss is defined as follows:
\begin{align}
\mathcal{L}_{\text{MI}} &= \hat{I}(Z_\text{emo},Z_\text{spk}) 
\end{align}
where \(\hat{I}\) is the estimate of the mutual information \cite{mackay2003information}.

\subsubsection{Identification loss} 
To enhance the modeling of the latent space, we integrate emotion and speaker identity losses as supervisory signals into two dedicated branches of the discriminator—one focusing on emotion and the other on speaker identity. By incorporating both emotion and speaker labels, the framework effectively learns to capture and distinguish these characteristics. Consequently, the emotion classifier categorizes the emotional content into five distinct domains, while the speaker discriminator determines whether a given speech feature corresponds to the same speaker.
\begin{align}
        \mathcal{L}_{\text{spk}} &= \mathbb{E}_{X,Y_\text{spk}}[CE( \psi_{spk}(Z_\text{spk});Y_\text{spk} ))] 
\end{align}
\begin{equation}
    \begin{aligned}
        \mathcal{L}_{\text{emo}} &= \mathbb{E}_{X,Y_\text{emo}}[CE(\psi_{emo}(Z_\text{emo});Y_\text{emo} )]
    \end{aligned}
    \label{}
\end{equation}
where $CE(\cdot;\cdot)$ is the cross entropy, and $\psi_{spk},\psi_{emo}$ are linear classifiers for speaker and emotional style classification. $Y_\text{spk}, Y_\text{emo}$ are the corresponding ground truth for speakers and emotional styles.

%

\subsubsection{Full objective}
The final loss is a combined sum:
\begin{align}
  \mathcal{L} = \mathcal{L}_{\text{REC}} + \mathcal{L}_{\text{KL}} +  \mathcal{L}_{\text{spk}} + \mathcal{L}_{\text{emo}} + \mathcal{L}_{\text{MI}}
\end{align}

\begin{figure}[t]
\begin{minipage}[b]{1.0\linewidth}
  \centering
  \centerline{\includegraphics[width=8.0cm]{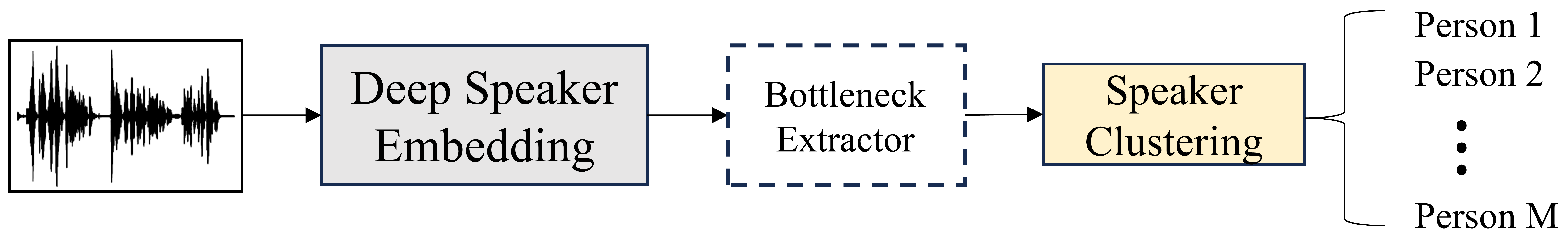}}
\end{minipage}
\caption{Speaker Clustering Process}
\label{fig4}
\end{figure}

\begin{table*}[t]
\centering
\caption{Comparison of the performance of different methods for speaker clustering on the ESD and IEMOCAP databases. Results are presented in terms of the performance of KM/SC/AC.\textbf{Bold} indicates the highest value, and \underline{underline} denotes the second highest value.}
\scalebox{1.0}{
\setlength{\tabcolsep}{12pt}
\begin{tabular}{l|c|c|c|c|c|c}
\hline
\multirow{2}{*}{}  & \multicolumn{3}{|c}{ESD} & \multicolumn{3}{|c}{IEMOCAP} \\
\cline{2-7}
& NMI$\uparrow$ & ARI$\uparrow$ &  Silhouette$\uparrow$ & NMI$\uparrow$ & ARI$\uparrow$  & Silhouette$\uparrow$ \\
\hline
\multicolumn{7}{c}{d-vector} \\
\hline

Baseline & 0.81/0.93/0.90 & 0.73/0.87/0.85 & 0.13/0.14/0.13 & 0.29/0.42/0.31 & 0.16/0.24/0.18 & 0.07/0.07/0.04\\
+ FT & 0.87/\textbf{0.97}/\textbf{0.97} & 0.74/\textbf{0.98}/\textbf{0.98} & 0.53/0.57/0.57 & 0.74/\underline{0.78}/0.75 & 0.65/\underline{0.73}/0.67 & 0.34/0.34/0.31 \\
+ VAE & 0.39/0.59/0.50 & 0.28/0.44/0.37 & 0.09/0.08/0.07 & 0.26/0.29/0.24 & 0.10/0.13/0.09 & 0.08/0.08/0.08 \\
+ ${\beta}$-VAE & 0.31/0.40/0.58 & 0.30/0.41/0.36 & 0.02/0.01/0.01 & 0.27/0.29/0.27 & 0.28/0.20/0.24 & 0.02/0.01/0.01\\
+ InfoVAE & 0.55/0.63/0.56 & 0.36/0.47/0.36 & 0.27/0.24/0.24 & 0.29/0.28/0.29 & 0.13/0.12/0.12 & 0.25/0.25/0.24 \\
\hline
+ DTG-VAE & \textbf{0.96}/\underline{0.96}/\underline{0.96} & \textbf{0.97}/\underline{0.97}/\underline{0.96} & \textbf{0.80}/\textbf{0.80}/\textbf{0.79} & \underline{0.77}/\textbf{0.79}/\textbf{0.80} & \underline{0.73}/\textbf{0.76}/\textbf{0.77} & \textbf{0.67}/\textbf{0.67}/\textbf{0.64}\\
- $L_{emo}$ & \underline{0.94}/0.92/0.95 & \underline{0.95}/0.92/0.95 & 0.73/0.70/\underline{0.72} & \textbf{0.78}/\textbf{0.79}/0.76 & \textbf{0.75}/\textbf{0.76}/0.70 & 0.54/0.54/0.53\\
- $L_{spk}$  & 0.25/0.21/0.26 & 0.13/0.09/0.12 & 0.27/0.20/0.22 & 0.08/0.08/0.07 & 0.02/0.02/0.01 & 0.26/0.18/0.24\\
- $L_{MI}$ & \textbf{0.96}/0.94/\underline{0.96} & \textbf{0.97}/0.92/\underline{0.96} & \underline{0.79}/\underline{0.74}/\textbf{0.79} & 0.73/0.76/\underline{0.77} & 0.70/0.71/\underline{0.71} & \underline{0.58}/\underline{0.58}/\underline{0.57}\\
\hline

\multicolumn{7}{c}{r-vector} \\
\hline
Baseline & 0.90/0.78/\textbf{0.99} & 0.87/0.43/\textbf{0.99} & 0.21/0.08/0.21 & 0.79/0.87/0.83 & 0.70/0.85/0.74 & 0.10/0.10/0.10 \\
+ FT &  \underline{0.98}/\underline{0.95}/\textbf{0.99} & 0.94/\underline{0.86}/\underline{0.96} & \underline{0.77}/\textbf{0.62}/\underline{0.77} & 0.88/\underline{0.89}/0.83 & 0.86/\underline{0.87}/0.78 & 0.55/0.54/0.53\\
+ VAE & 0.94/0.84/\textbf{0.99} & 0.86/0.66/\textbf{0.99} & 0.17/0.19/0.19 & 0.67/0.82/0.69 & 0.53/0.74/0.53 & 0.10/0.12/0.11 \\
+ ${\beta}$-VAE & 0.25/0.51/0.25 & 0.19/0.36/0.18 & 0.37/-0.05/0.39 & 0.71/0.70/0.60 & 0.43/0.40/0.37 & 0.01/-0.02/0.01 \\
+ InfoVAE & \textbf{0.99}/0.86/\textbf{0.99} & \textbf{0.99}/0.66/\textbf{0.99} & 0.26/0.14/0.26 & 0.66/0.77/0.70 & 0.53/0.66/0.51 & 0.11/0.12/0.11\\
\hline
+ DTG-VAE & \textbf{0.99}/\textbf{0.97}/\textbf{0.99} & \textbf{0.99}/\textbf{0.96}/\textbf{0.99} & \textbf{0.88}/\underline{0.40}/\textbf{0.88} & \textbf{0.92}/\textbf{0.92}/\textbf{0.91} & \textbf{0.91}/\textbf{0.90}/\textbf{0.89} & \textbf{0.73}/\textbf{0.73}/\textbf{0.72}\\
- $L_{emo}$ & \underline{0.98}/0.79/\underline{0.96} & \underline{0.98}/0.79/\textbf{0.99} & 0.27/0.27/0.27 & \underline{0.90}/0.85/\underline{0.89} & \textbf{0.91}/\textbf{0.90}/\underline{0.86} & 0.51/0.51/0.50\\
- $L_{spk}$ & 0.24/0.25/0.24 & 0.15/0.11/0.14 & 0.23/0.18/0.18 & 0.74/0.78/0.72 & 0.66/0.69/0.61 & 0.21/0.20/0.20\\
- $L_{MI}$ & 0.89/0.81/0.90 & 0.89/0.56/0.91 & 0.72/0.25/0.72 & 0.88/0.87/0.82 & \underline{0.87}/\underline{0.87}/0.81 & \underline{0.69}/\underline{0.69}/\underline{0.65}\\
\hline

\multicolumn{7}{c}{ECAPA-TDNN} \\
\hline
Baseline &  0.95/0.93/\textbf{0.99} & 0.87/0.82/\textbf{0.99} & 0.16/0.18/0.19 & 0.79/0.82/0.76 & 0.67/\underline{0.78}/0.62 & 0.09/0.09/0.08 \\
+ FT & \textbf{0.99}/0.89/\textbf{0.99} & \textbf{0.99}/0.68/\textbf{0.99} & 0.49/0.28/0.49 & 0.44/0.65/0.50 & 0.36/0.52/0.37 & 0.05/0.05/0.03 \\
+ VAE & 0.92/\underline{0.99}/0.97 & 0.85/\textbf{0.99}/\underline{0.97} & 0.16/0.17/0.17 & 0.60/0.70/0.62 & 0.42/0.50/0.32 & 0.09/0.10/0.08\\
+ ${\beta}$-VAE & 0.95/0.83/\textbf{0.99} & 0.87/0.55/\textbf{0.99} & 0.37/0.24/0.40 & 0.71/0.72/0.73 & 0.60/0.52/0.63 & \underline{0.40}/\underline{0.38}/\underline{0.39}\\
+ InfoVAE & \underline{0.97}/\underline{0.99}/\underline{0.98} & \underline{0.97}/\textbf{0.99}/\textbf{0.99} & 0.16/0.16/0.16 & 0.65/0.75/0.66 & 0.51/0.55/0.43 & 0.10/0.11/0.10  \\
\hline
+ DTG-VAE & \textbf{0.99}/\textbf{1.00}/\textbf{0.99} & \textbf{0.99}/\textbf{0.99}/\textbf{0.99} & \textbf{0.74}/\textbf{0.74}/\textbf{0.75} & \textbf{0.88}/\textbf{0.87}/\textbf{0.82} & \textbf{0.84}/0.75/\underline{0.67} & \textbf{0.48}/\textbf{0.47}/\textbf{0.44}\\
- $L_{emo}$ & \textbf{0.99}/0.95/\textbf{0.99} & \textbf{0.99}/0.86/\textbf{0.99} & \underline{0.59}/\underline{0.47}/\underline{0.59} & \underline{0.84}/\underline{0.83}/\underline{0.78} & \underline{0.82}/\textbf{0.80}/\textbf{0.70} & 0.36/0.36/0.33\\
- $L_{spk}$ & 0.89/0.95/0.95 & 0.67/\underline{0.87}/0.95 & 0.05/0.08/0.08 & 0.61/0.68/0.62 & 0.46/0.41/0.33 & 0.06/0.06/0.05\\
- $L_{MI}$ & 0.93/\underline{0.99}/\textbf{0.99} & 0.86/\textbf{0.99}/\textbf{0.99} & 0.31/0.34/0.34 & 0.76/0.77/0.75 & 0.65/0.65/0.63 & 0.26/0.30/0.28\\
\hline

\end{tabular}
}
\label{table2}
\end{table*}

\section{Experiment}
\label{sec:Experiment}
\subsection{Dataset}
We continue to utilize the IEMOCAP \cite{busso2008iemocap} and ESD \cite{zhou2022emotional} datasets, focusing on five emotions: anger, surprise, happiness, neutral, and sadness. Each dataset is divided randomly in an 80\%/10\%/10\% ratio for training, validation, and testing. We calculate the mean performance across 10-fold cross-validation.

\subsection{Model Architecture \& Training}
In our experiment, our $E_{share}$ encoder consists of 3 fully-connected layers with a ReLU activation function that projects the deep speaker embedding into 256 dimensions. The speaker encoder and emotion encoder are composed of 3 fully-connected layers. Each Dense layer is followed by Layer Normalization layer. This output vector is then projected into the mean and logarithm vectors of the distribution parameters of the speaker and emotion posterior distribution. The dimension of the speaker and emotion posterior distribution is set to 256. The reparameterization trick allows us to sample from the speaker and emotion posterior distribution, which yields a 256-dimension speaker and emotion latent vector. The decoder aims to form the original deep speaker embedding. We concatenate two latent representations on the feature axis. The speaker latent vector is duplicated to match the length of the emotion latent vector. Our decoder consists of 2 fully-connected dense layers with a ReLU activation function. Our models undergo 400 epochs of training with a learning rate of 1e-4, stopping based on the validation accuracy using the Adam optimizer with a batch size of 32. We repeat each supervised training five times with different initialization seeds and measure NMI, ARI, and Silhouette during the evaluation. 

\subsection{Results and Discussion}

After our DTG-VAE model has been fully trained, we employ a bottleneck extractor to obtain more accurate speaker embeddings for clustering, as shown in Figure \ref{fig3} and Figure \ref{fig4}. Table \ref{table2} presents our speaker clustering results using the KM/SC/AC methods. Although r-vector and ECAPA-TDNN perform better ability of capturing speaker identity than d-vector in baseline, they both perform limited in silhouette score. DTG-VAE significantly improves speaker clustering performance across three clustering algorithms compared to directly fine-tuning deep speaker embeddings and other VAE methods in most cases, especially in the Silhouette metric. The ablation study reveals that the speaker encoder module is the most critical component for disentanglement within our method, while the emotion encoder contributes to extracting more accurate speaker embeddings. Mutual information loss effectively disentangles the emotional and speaker components, thereby continuously enhancing the accuracy of speaker clustering. DTG-VAE leverages pre-trained models directly, resulting in significant time and cost savings. Currently, we have only tested on two datasets with ten speakers each and our method requires both emotion-labeled and speaker-labeled data. 

\section{Conclusion}
\label{sec:Conclusion}
We propose a novel, potentially disentangled model structure named DTG-VAE that facilitates the extraction of more effective and accurate speaker embeddings for speaker clustering tasks.  Our model architecture achieves superior performance compared to directly fine-tuning a pre-trained model. Additionally, our approach can be adapted to emotion-unlabeled datasets. In future work, we will investigate the applicability of our framework to data without labeled emotions.

\begingroup  
\bibliographystyle{IEEEtran} 
\bibliography{conference_101719} 
\endgroup

\end{document}